\documentclass[journal]{IEEEtran}
\usepackage{cite}

\ifCLASSINFOpdf
  \usepackage[pdftex]{graphicx}
  \graphicspath{{fig/}}
\else
  \usepackage[dvips]{graphicx}
  \graphicspath{{fig/}}
\fi
\usepackage{multirow}

\usepackage[cmex10]{amsmath}
\usepackage{multirow}

\ifCLASSOPTIONcompsoc
  \usepackage[caption=false,font=normalsize,labelfont=sf,textfont=sf]{subfig}
\else
  \usepackage[caption=false,font=footnotesize]{subfig}
\fi
\begin{document}
%
\title{Deep Learning Based NLOS Identification\\with Commodity WLAN Devices}
%
%
%

\author{Jeong-Sik Choi, Woong-Hee Lee, Jae-Hyun Lee, Jong-Ho Lee,~\IEEEmembership{Member,~IEEE},\\ and Seong-Cheol Kim,~\IEEEmembership{Senior Member,~IEEE} 
\thanks{Copyright (c) 2015 IEEE. Personal use of this material is permitted. However, permission to use this material for any other purposes must be obtained from the IEEE by sending a request to pubs-permissions@ieee.org.}
\thanks{This work was supported by the National Research Foundation of Korea (NRF) grant funded by the Korea government (MSIP) (No. NRF-2015R1A2A2A03008195).}
\thanks{Jeong-Sik Choi is with Intel Labs, Intel Corporation, Santa Clara, CA, USA (e-mail: jeongsik.choi@intel.com).}
\thanks{Woong-Hee Lee is with the Advanced Standard and Development Labs, LG Electronics, Seoul, Korea (e-mail: wh.lee@lge.com).}
\thanks{Jae-Hyun Lee and Seong-Cheol Kim are with the Department of Electrical Engineering and INMC, Seoul National University, Seoul, Korea (e-mail: \{sacrifice57, sckim\}@maxwell.snu.ac.kr).}
\thanks{Jong-Ho Lee is with the Department of Electrical Engineering, Gachon University, Seongnam, Gyeonggi, Korea (e-mail: jongho.lee@gachon.ac.kr).}}

%
%

\markboth{Accepted for publication in IEEE Transactions on Vehicular Technology}{}
%



\maketitle

\begin{abstract}
Identifying line-of-sight (LOS) and non-LOS (NLOS) channel conditions can improve the performance of many wireless applications, such as signal strength-based localization algorithms.
For this purpose, channel state information (CSI) obtained by commodity IEEE 802.11n devices can be used, because it contains information about channel impulse response (CIR).
However, because of the limited sampling rate of the devices, a high-resolution CIR is not available, and it is difficult to detect the existence of an LOS path from a single CSI measurement, but it can be inferred from the variation pattern of CSI over time.
To this end, we propose a recurrent neural network (RNN) model, which takes a series of CSI to identify the corresponding channel condition.
We collect numerous measurement data under an indoor office environment, train the proposed RNN model, and compare the performance with those of existing schemes that use handcrafted features.
The proposed method efficiently learns a non-linear relationship between input and output, and thus, yields high accuracy even for data obtained in a very short period.
\end{abstract}

\begin{IEEEkeywords}
Line-of-sight identification, indoor localization, channel state information, recurrent neural network, long short-term memory.
\end{IEEEkeywords}

%
\IEEEpeerreviewmaketitle

\section{Introduction}
%
%
%
%
\IEEEPARstart{T}{he} existence of a line-of-sight (LOS) path between a transmitter and a receiver generally determines the property of a wireless link because it travels the shortest distance without experiencing additional losses due to reflection, diffraction, and scattering.
Such a direct path usually dominates other multi-path components, and consequently, lowers path loss.
Because of the different characteristics across LOS and non-LOS (NLOS) conditions, identifying these two channel conditions and applying an appropriate strategy to each condition can improve the performance of many wireless applications, for instance, signal strength-based localization algorithms~\cite{6853330, 7736965}.

To achieve this objective, several strategies for identifying channel conditions have been explored in the literature, with an especially large portion of effort dedicated to ultra-wide-band (UWB) systems \cite{Guvenc:2008:NIW:1376536.1340500, 4212756, 4224541, 4407489, 5555901, 4907427}.
Because of the extremely wide frequency band, a high-resolution channel impulse response (CIR) can be captured at the receiver.
For this reason, multi-path components are easily separated in the time axis, and the existence of a dominant path in the multi-path profile is easily recognizable through various features such as time of arrival, delay spread, and the statistics of the received signal strength.

However, UWB functionality is rarely included in off-the-shelf wireless devices these days.
As a result, many wireless localization techniques are being developed on the basis of more widespread wireless communication systems, such as cellular or wireless local area network (WLAN) systems~\cite{Youssef:2005:HWL:1067170.1067193, 1458284, monalisa, 7438932}.
These systems widely adopt orthogonal frequency division multiplexing (OFDM) techniques to enhance the spectral efficiency and typically occupy narrow frequency bands compared to UWB systems.
This is one of the major reasons that the identification of LOS and NLOS channel conditions is still necessary for wireless communication systems operating across only a few tens of MHz bandwidth.


Because the time resolution of a multi-path profile is normally limited by the sampling rate, which is relevant to the system bandwidth, a high-resolution CIR is not available in commodity devices for cellular and WLAN systems.
Therefore, it is difficult to decide channel conditions by observing a single snapshot of signal transmission.
To overcome this problem, authors in \cite{6963485} tried to infer whether there exists a dominant component in a multi-path propagation channel by monitoring the fluctuation in signal strength over time.
For instance, one of the most important clues in identifying channel conditions, the Rician $K$ factor, can be estimated from the shape of signal strength distribution~\cite{913150, Koay2006317}.

However, the received signal strength is a representative value obtained by averaging the strength of every available sub-carrier in OFDM systems.
Therefore, we can have more rich information to identify channel conditions if the frequency response of each sub-carrier, which is also known as channel state information (CSI) in the IEEE 802.11 standard~\cite{IEEE802.11}, is provided.
For instance, if the amplitude of CSI tends to be flat over several consecutive sub-carriers, it means that there exists a strong dominant component in CIR.

In this context, several studies have proposed handcrafted features for identifying channel conditions by monitoring a series of CSI over time~\cite{6646249, 6848217,7130677,7218588}.
The skewness of dominant path power and the kurtosis of frequency diversity variation are proposed in~\cite{7130677}.
These features focus on different time-varying characteristics of each channel condition, and thus, are more suitable to use in a mobile scenario.
Another recently discovered feature is the variation in phase, extracted from CSI~\cite{7218588}.
Unlike the previous features, this feature is valid for a stationary scenario, where the number of multi-path components remains unchanged during consecutive transmissions.
Some studies have even tried to improve the identification accuracy by exploiting multiple antennas~\cite{6646249, 7218588}.

Even though the aforementioned handcrafted features yield high accuracy under certain scenarios, it is uncertain that they extract every useful information from CSI.
In addition, features introduced in~\cite{7130677, 7218588} are based only on CSI, but we believe that an upper layer information, the received signal strength indicator (RSSI), can also play an assistive role in channel condition identification.
For instance, if RSSI is too high or too low in a particular environment, it is very likely that the environment is LOS or NLOS conditions, respectively.

In this situation, a deep learning approach can play an important role, because it efficiently learns the complex non-linear relationship between raw input and output data through training process.
Moreover, if the input data contain cross-layer information, the neural network learns how to combine this information appropriately.
As previous studies in~\cite{6963485, 6646249, 6848217,7130677,7218588} have claimed, a single snapshot of signal transmission carries insufficient information to identify channel conditions, especially when the system bandwidth is narrow.
Therefore, even a deep learning approach is not able to make an accurate decision from a limited number of observations.

For this reason, we also try to identify the channel condition from a series of CSI, and consider a recurrent neural network (RNN) model, which can effectively deal with sequential input data~\cite{RNN}.
As the term recurrent implies, the RNN model takes not only the current input data but also several previous input data.
In other words, it has a memory that can capture the variation in input data.
In addition, more advanced memory units, such as the long short-term memory (LSTM)~\cite{LSTM1997} or gated recurrent unit (GRU)~\cite{GRU}, have been proposed in the literature. They efficiently control the flow of information using special units called gates.
Before using the RNN model, parameters in the model should be determined appropriately through training data.
This paper makes the following contributions:

\begin{enumerate}
\item{To consider a series of CSI data efficiently, we propose an RNN structure consisting of an LSTM block for identifying channel conditions. From the CSI measured by each signal transmission, we create appropriate input vectors for the proposed RNN model and train the model using the measurement data.
}
\item{To obtain training and test data for a practical environment, we carry out excessive measurement campaigns under an indoor office environment with commodity WLAN devices. We collect numerous data with respect to various locations of the transmitter and the receiver. Especially, we consider two types of NLOS conditions, where the direct path is shadowed owing to building structures or a human body.
}
\item{We verify the performance of the proposed RNN model with conventional schemes. The proposed method produces high accuracy even for the data observed in a very short time. Therefore, it significantly reduces the time required for identifying channel conditions, and thus, saves energy consumption for both signal transmission and reception.
    Moreover, such an instant decision ability is desirable to capture a quick transition between the channel conditions.}
\end{enumerate}

The remainder of the paper is organized as follows.
The problem formulation and channel model are explained in Section~II.
The measurement environment and scenario are introduced in Section~III, and the proposed neural network model is expressed in Section~IV. The performance of the proposed and conventional methods is verified in Section~V, and the paper is concluded in Section VI.

\section{Preliminaries}

\subsection{Problem Formulation}

We consider a WLAN system where a receiver estimates the frequency responses of ODFM sub-carriers by receiving signal transmissions from a transmitter.
Since the frequency response of a sub-carrier is called CSI in the IEEE 802.11 standard~\cite{IEEE802.11}, we use these two terms interchangeably.
For simplicity, the transmitter and receiver can be considered as a WLAN access point (AP) and a terminal station, respectively.
It is assumed that the AP periodically broadcasts signals into air, such as beacon transmission.
According to this assumption, the receiver is even able to predict the channel condition from unconnected APs in the vicinity.

The receiver can measure a bunch of CSI per signal reception using the commodity WLAN devices.
By receiving several transmitted signals from the transmitter, the receiver acquires a series of CSI, and then, predicts whether the current propagation channel is LOS or NLOS conditions.
Clearly, the LOS condition indicates that there is no obstacle between the transmitter and receiver.
Because the signal strength between the transmitter and receiver is also affected by human bodies~\cite{5208178, 6778037, 7368073}, it is natural to assume that the NLOS condition includes the case where the direct path is shadowed by a human body.

\subsection{Channel Model}

Because of the sampling process, the receiver captures a discrete version of the CIR, which is expressed by
$\mathbf{h}=[h(0), h(1), ..., h(M-1)]^T$, where $M$ represents the number of multi-path taps.
The sampling interval is defined by $T_s$, and the receiver cannot discriminate multiple rays arriving within the same sampling interval.
Instead, it recognizes such multiple rays as a single tap, whose coefficient is generated by the superposition of the rays.
For simplicity, channel variations within one block of data are considered negligible.
By applying the Fourier transform, the frequency response of the $n$-th sub-carrier is expressed by
\begin{equation}\label{2A1}
H(f_n)=\sum_{m=0}^{M-1} h(m)e^{-j 2 \pi f_n m T_s},
\end{equation}
where $f_n$ is the frequency of the $n$-th sub-carrier.
The spacing between any of the two consecutive sub-carriers is expressed by $\Delta f=\frac{1}{NT_s}$, where $N$ is the DFT size.

For channel estimation, the transmitter transmits known symbols through selected sub-carriers and the receiver can obtain the frequency-domain channel coefficients by receiving this transmission.
The CSI of the $n$-th sub-carrier is estimated  at the receiver  as follows:
\begin{equation} \label{2A2}
\hat{H}(f_n)=H(f_n)+v_n,\ n\in\mathcal S,
\end{equation}
where $v_n$ is a complex Gaussian noise for the $n$-th sub-carrier with zero mean and variance $N_0$, and $\mathcal S$ represents the set of sub-carriers used in signal transmission.

According to the IEEE 802.11n standard, not all sub-carriers are involved in signal transmission.
For instance, the DFT size of the 802.11n system operating with 20~MHz bandwidth is 64, but only 56 sub-carriers close to the center frequency are used.
For this reason, the receiver can obtain the set of CSI with respect to sub-carriers that are actually involved in signal transmission.
For this reason, $\mathcal S$ can be represented by $\mathcal S=\{-28, -27, ..., -1, 1, ..., 27, 28\}$.
Some commodity WLAN devices provide a representative CSI for a group of consecutive sub-carriers, and thus, only 30 or 16 CSI may be available for the 20~MHz bandwidth system.

\section{Measurement Campaigns}

\begin{figure}[!t]
\centering
\includegraphics[width=0.48\textwidth]{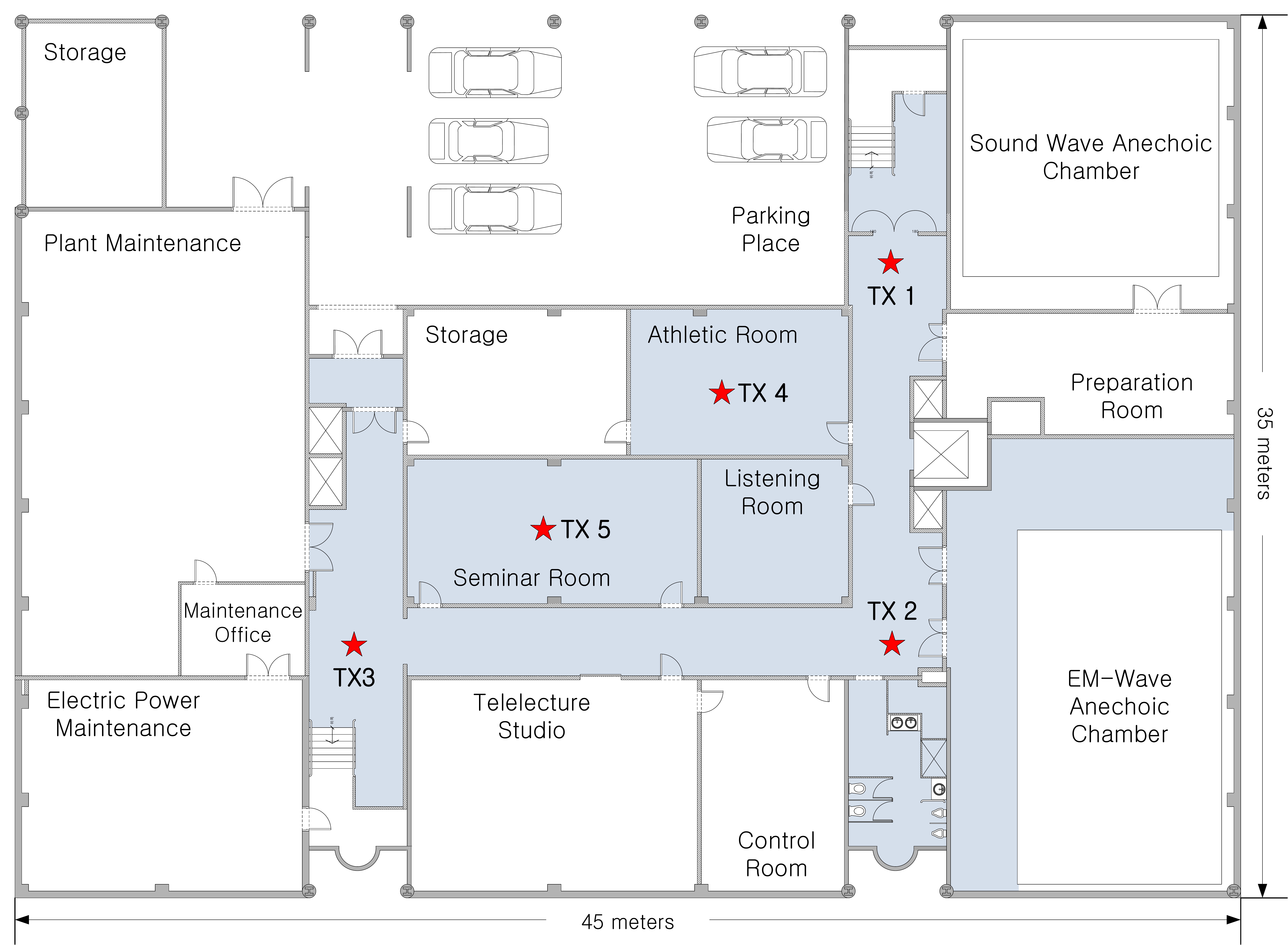}
\caption{Layout of the measurement site sized 45$\times$35~m$^2$. Three transmitters are located in corridors and two are located in rooms. The receiver can move in the highlighted areas.}
\label{figLayout}
\end{figure}

To obtain data for real environments, we conducted measurement campaigns at a laboratory building in Seoul National University (SNU), which can be considered as a typical indoor office environment.
Fig.~\ref{figLayout} illustrates the layout of the measurement site and five locations for the transmitter.
For each transmitter location, a person holding the receiver moves around in all accessible areas highlighted in the figure to collect measurement data.
The speed of the receiver is around 0.5~m/s, and the height of both the transmitter and receiver is fixed at 1.2~m.

We use two laptops installed with Qualcomm Atheros devices (i.e., AR9462 and AR9480) as the transmitter and receiver.
Since these network interface cards support two and three antennas, respectively, we can simultaneously acquire up to six spatial CSI sets at each signal transmission.
The Atheros CSI tool running on a Linux environment at the receiver provides non-grouping and non-compressed information~\cite{atheros}.
In other words, it provides CSI data for all sub-carriers involved in signal transmission (i.e., 56 sub-carriers for 20~MHz bandwidth) and each CSI is provided by 10-bit resolution in both real and imaginary parts.

The carrier frequency is 2.462~GHz (i.e., WLAN channel~11) and the bandwidth of the system is 20~MHz.
The transmitter sends a packet every 10~ms and the receiver records CSI with the information of the current channel condition.
As we mentioned in Section II, the LOS condition is recorded if both the transmitter and receiver can see each other.
Both the transmitter and receiver are equipped with isotropic antennas with 5~dBi gain.
The receiver measured 101,197 packet transmissions under the LOS condition, 103,547 under the NLOS condition due to human body shadowing, and 227,818 under the NLOS condition due to a building structure (e.g., shadowing because of wall and door).
The total time purely consumed for data collection was therefore 4300~s.
Note that we can acquire six sets of CSI data from each packet transmission due to the multiple antenna configuration.

\begin{figure}[!t]
\centering
\includegraphics[width=0.48\textwidth]{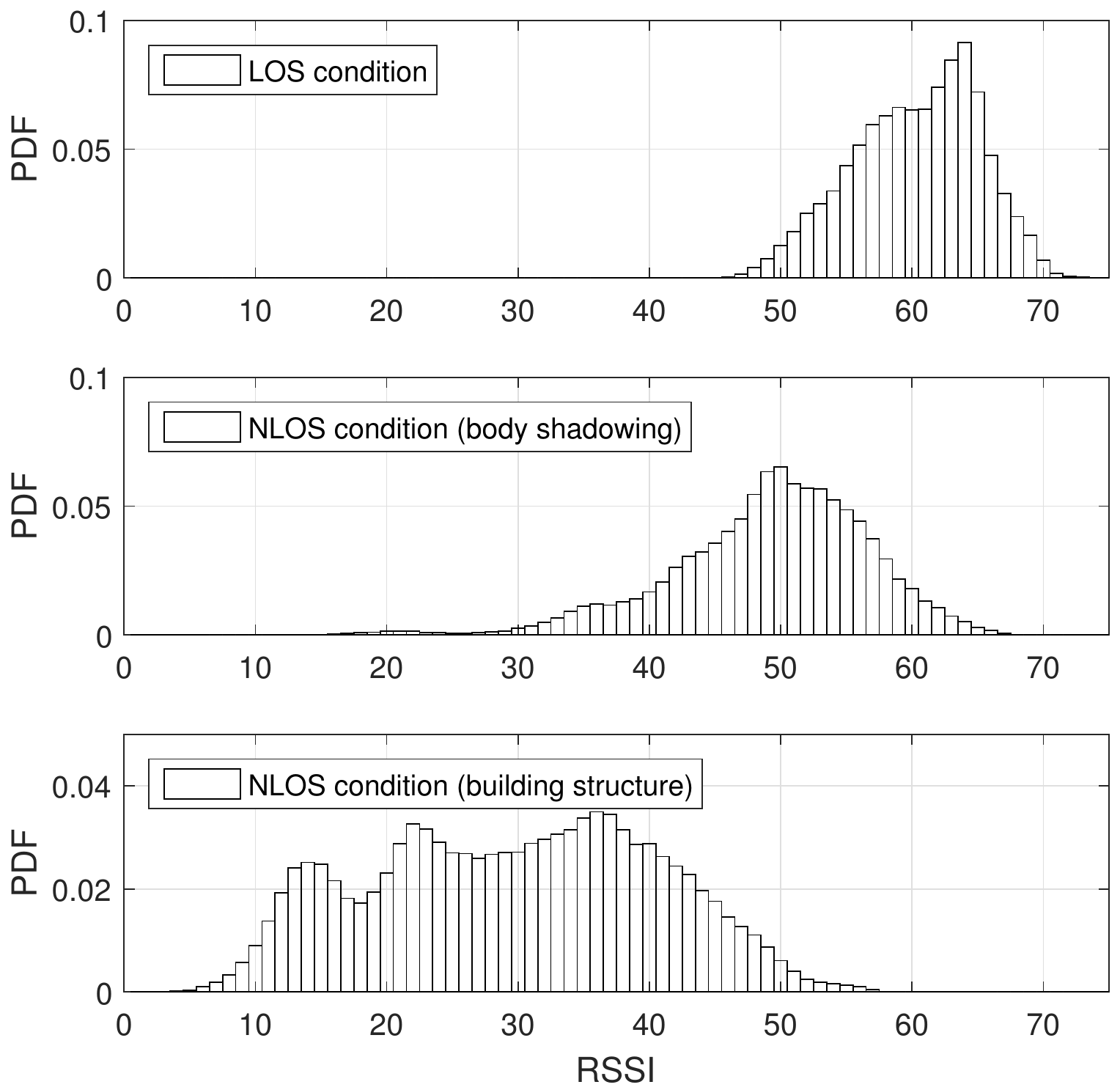}
\caption{Distribution of RSSI for LOS and NLOS conditions due to human body and building structure.}
\label{figRssiPDF}
\end{figure}

In addition to CSI, the Atheros CSI tool also provides upper layer information, RSSI, for each transmission.
Fig.~\ref{figRssiPDF} depicts the distributions of RSSI measured under LOS and NLOS conditions.
In this figure, we separate NLOS data into two cases according to the type of obstacle.
Since the receiver is equipped with three antennas, three RSSIs are measured in each signal transmission, and we simply take their average for representing this figure.
As we can see, the signal strength under the LOS condition is concentrated in a high value, where every RSSI is greater than 40.
On the other hand, RSSI for the NLOS condition is distributed over a wide range.

\section{Neural Network Model for Identifying Channel Condition}

To consider a series of CSI data, we use an RNN structure in this study.
The first step is to create an appropriate input vector from the measurement data.
Because each CSI is a complex number, we consider real and imaginary parts separately.
Furthermore, we include RSSI in the input vector because it also provides important information.
Therefore, the input vector can be expressed by
\begin{equation} \label{4A1}
\mathbf{x}=[RSSI, H_R(f_{-28}), H_I(f_{-28}), ..., H_R(f_{28}), H_R(f_{28})]^T,
\end{equation}
where subscripts $R$ and $I$ indicate the real and imaginary parts of a complex number, respectively.
In this case, the dimension of the input vector is $D_x=113$, which can be reduced if fewer CSIs are available.

\begin{figure}[tp]
\centering
    \subfloat[]{
        \includegraphics[width=0.43\textwidth]{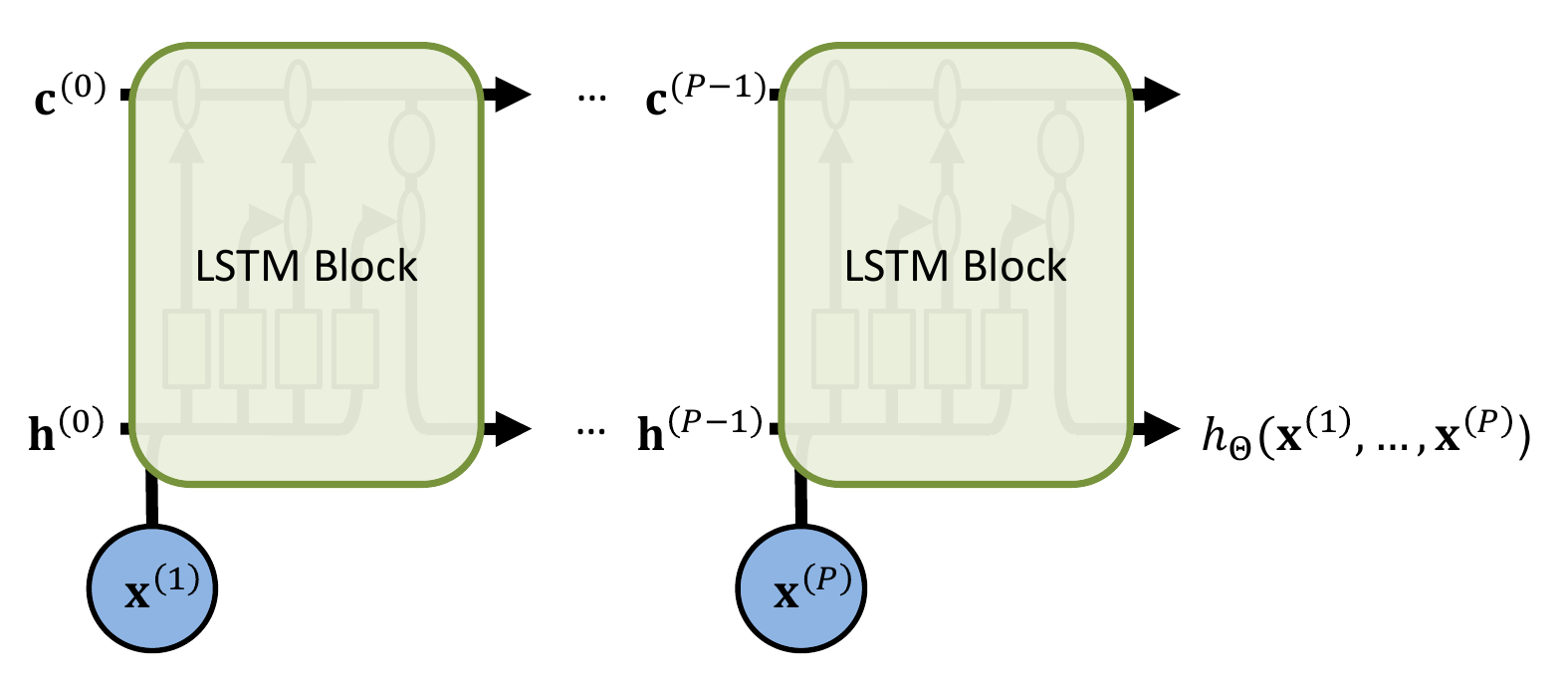}}
    \vfil
    \subfloat[]{
        \includegraphics[width=0.45\textwidth]{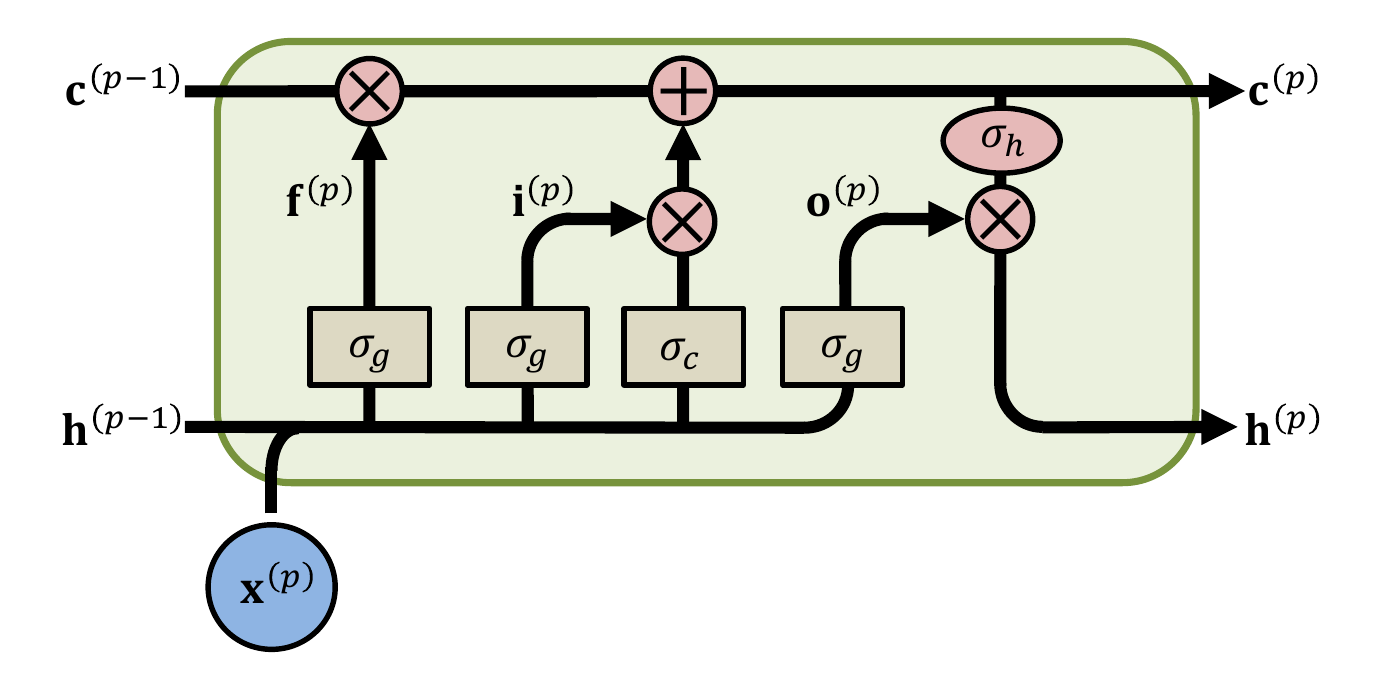}}
    \caption{RNN structure with the LSTM block: (a) overall structure, (b) detailed structure of the LSTM block at the $p$-th time step.}
    \label{figRNN}
\end{figure}

For fair performance comparison with existing schemes \cite{7130677, 7218588}, which use $P$ sets of CSI measurements, we consider an RNN model taking the same number of CSI sets to make a decision.
Fig.~\ref{figRNN} illustrates the RNN model with the LSTM block, where $\mathbf{x}^{(p)}$ and $\mathbf{h}^{(p)}$ represent input vector and hidden layer at the $p$-th time step, respectively.
We assume that the dimension of the hidden layer is $D_h$.
In addition to the hidden layer, the LSTM block has one more unit, called cell state $\mathbf{c}^{(p)}$, which controls the flow of information using three gates: forget, input, and output gates.

With respect to the input vector, each gate vector is calculated as follows:
{\setlength\arraycolsep{1pt}
\begin {eqnarray} \label{4A2}
&&\mathbf{f}^{(p)}=\sigma_g(\mathbf{W}_f\mathbf{x}^{(p)}+\mathbf{U}_f\mathbf{h}^{(p-1)}+\mathbf{b}_f),\\ \nonumber
&&\mathbf{i}^{(p)}=\sigma_g(\mathbf{W}_i\mathbf{x}^{(p)}+\mathbf{U}_i\mathbf{h}^{(p-1)}+\mathbf{b}_i),\\ \nonumber
&&\mathbf{o}^{(p)}=\sigma_g(\mathbf{W}_o\mathbf{x}^{(p)}+\mathbf{U}_o\mathbf{h}^{(p-1)}+\mathbf{b}_o),
\end {eqnarray}}%
where every matrix denoted by $\mathbf W$ has $D_h \times D_x$ shape, those denoted by $\mathbf U$ have $D_h\times D_h$ shape, and the vectors denoted by $\mathbf b$ have $D_h\times 1$ shape.
Furthermore, $\sigma_g(\cdot)$ represents an element-wise activation function for gates, which is given by a sigmoid function in this study, i.e., $\sigma_g(z)=1/(1+e^{-z})$.
Both previous cell state and hidden layer of the LSTM block at the first time step are initialized as zero vectors.

Depending on the gate vectors, the cell state and hidden layer in the LSTM block are updated as
{\setlength\arraycolsep{1pt}
\begin {eqnarray} \label{4A3}
&\mathbf{c}^{(p)}=\mathbf{f}^{(p)}\circ \mathbf{c}^{(p-1)}+\mathbf{i}^{(p)}\circ \sigma_c(\mathbf{W}_c\mathbf{x}^{(p)}+\mathbf{U}_c \mathbf{h}^{(p-1)}+\mathbf{b}_c),\nonumber\\
&\mathbf{h}^{(p)}=\mathbf{o}^{(p)}\circ \sigma_h(\mathbf{c}^{(p)}),
\end {eqnarray}}%
where the operator $\circ$ represents the Hadamard product, and $\sigma_c(\cdot)$ and $\sigma_h(\cdot)$ are element-wise activation functions for the cell state and hidden layer, respectively.
In this study, we utilize the hyperbolic tangent (tanh) for these activation functions.
Finally, the output of the proposed model directly comes from the hidden layer at the last time step as follows:
\begin{equation} \label{4A4}
h_\Theta(\mathbf{x}_1, ..., \mathbf{x}_P)=\sigma(\mathbf{V}\mathbf{h}^{(P)}+b),
\end{equation}
where $\mathbf{V}$ is the $D_h$ dimensional row vector and $b$ is a bias constant.
The set $\Theta$ contains every parameter in the model, for example, elements in the matrices $\mathbf{U}$ and $\mathbf{W}$, vectors $\mathbf{b}$ and $\mathbf{V}$, and the constant $b$.

Each parameter is adjusted through training data. We denote the sequence of $P$ input data vectors as $\mathbf{X}=(\mathbf{x}_1, ..., \mathbf{x}_P)$ and the corresponding label as $y$.
For convenience, we assume that $y=1$ for LOS conditions and $y=0$ for NLOS conditions.
From the measurement data, we generate a set of input and output pairs $(\mathbf{X}, y)$ to train and verify the proposed RNN model.
We denote $\mathcal G$ as such a set, and every parameter in the model is adjusted in the direction of minimizing the following cost function:
\begin{equation}  \label{4A4}
J(\Theta)=-\frac{1}{|\mathcal G|}\sum_{g\in\mathcal G} C(g),
\end{equation}
where $|\cdot|$ is the number of elements in a set. $C(g)$ is the cost of the $g$-th input and output pair and it measures how accurately the RNN model produces the output compared to the ground truth data.
Among many choices of evaluating the cost, we use cross-entropy, which is express as
\begin{equation}
C(g)=y^{(g)}\log h_{\Theta}(\mathbf{X}^{(g)})+(1-y^{(g)})\log (1-h_{\Theta}(\mathbf{X}^{(g)})),
\end{equation}
where the superscript indicates the index of the input and output pair.


\begin{figure*}[!t]
\centering
\subfloat[]{\includegraphics[width=0.48\textwidth]{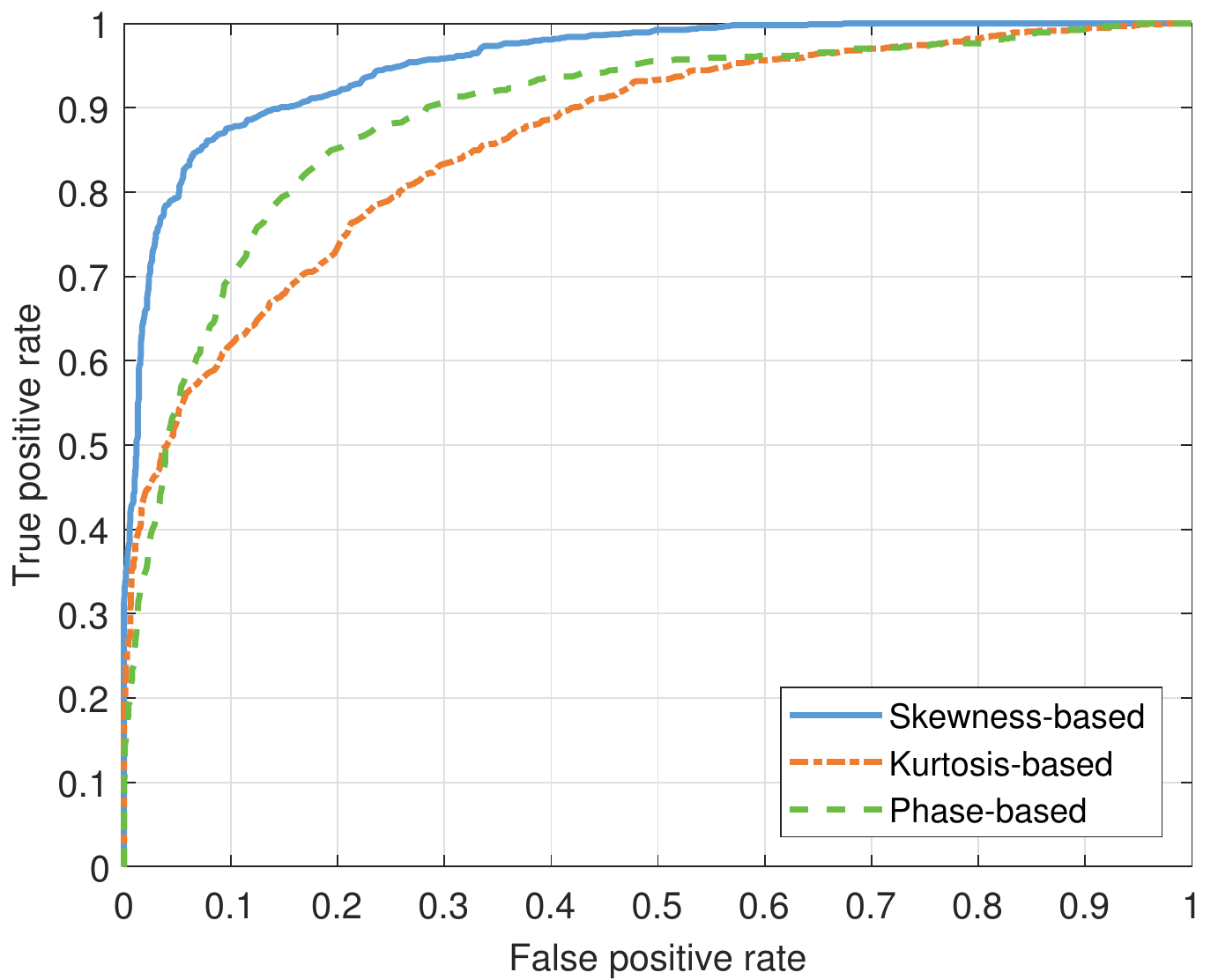}}
\hfill
\subfloat[]{\includegraphics[width=0.48\textwidth]{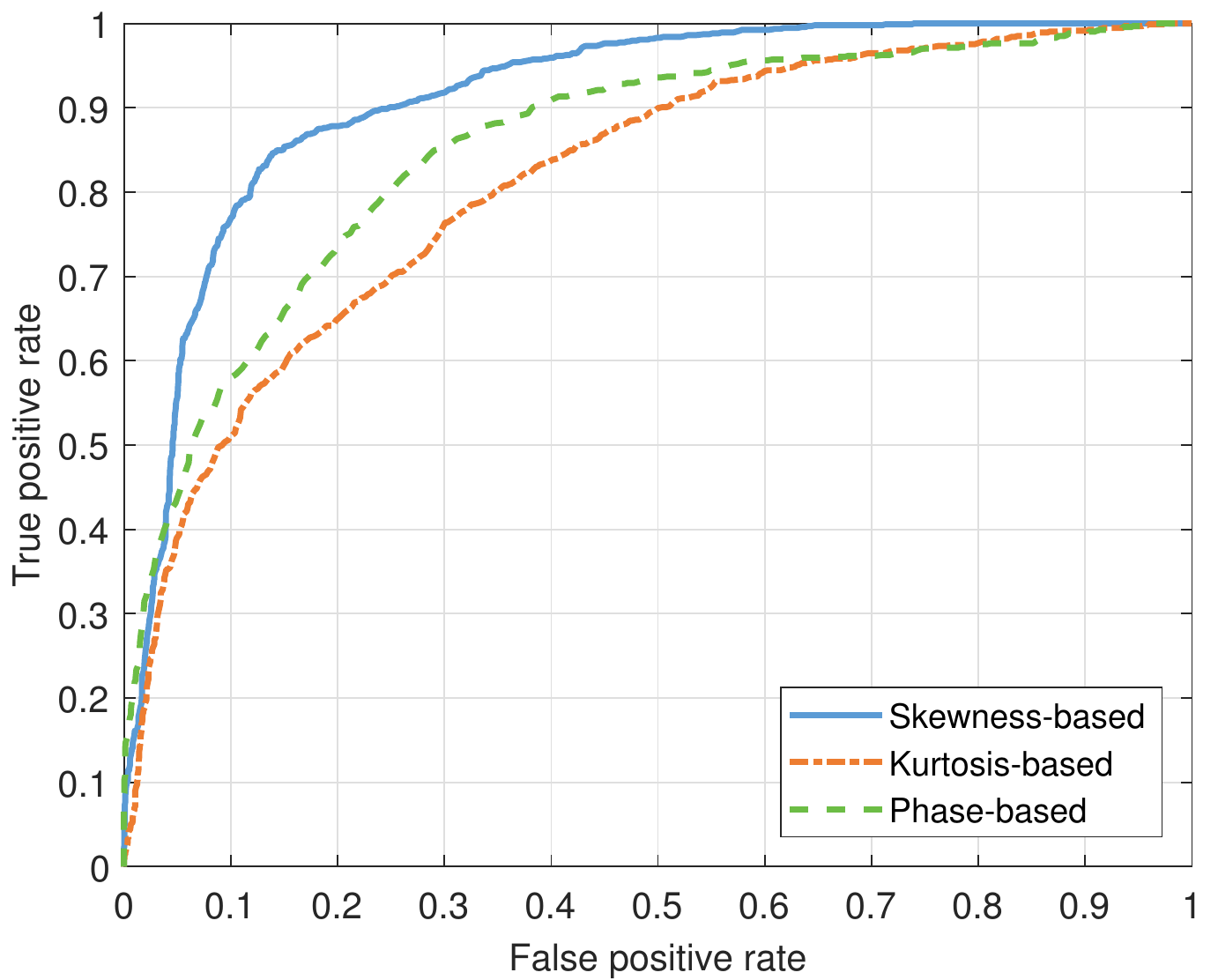}}
\caption{ROC curves of the conventional channel condition identification methods. Each handcrafted feature is obtained from $P=1000$ packet transmissions: (a) NLOS condition due to building structure only, (b) NLOS condition including body shadowing.}
\label{figRocConv}
\end{figure*}

Every parameter in the neural network is generally updated in an iterative manner using the gradient descent method.
It simultaneously updates every parameter at each iteration in the direction of the steepest descent, as follows:
\begin{equation}
\Theta_{n+1}=\Theta_n-\eta \nabla_{\Theta} J(\Theta),
\end{equation}
where $\nabla_\Theta$ is the gradient operator with respect to $\Theta$ and $\eta$ is the learning rate, which determines the step size.
Many variations of the gradient descent method have been studied in the literature, such as AdaGrad, AdaDelta, and Adam optimizers~\cite{AdaGrad, AdaDelta, Adam}.
These optimizers adaptively change the learning rate to precisely achieve the minimum cost.
Regardless of which optimizer is used, we need to take the gradient of the cost function at each iteration, and the back propagation (BP) algorithm quickly does that using matrix multiplications~\cite{backpropagation}.

Because of numerous parameters, the neural network can efficiently capture non-linearity between the input and output layers, but it has a high risk of over-fitting.
In other words, the neural network tends to remember all training data, instead of finding an efficient decision rule.
To prevent this, early stopping is used in this study.
We separate the set of input and output pairs, $\mathcal G$, into three non-overlapping sets: training set $\mathcal G_{tr}$, validation set $\mathcal G_{v}$, and test set $\mathcal G_{te}$.
Then, we train the proposed RNN model using the training set.
At each iteration, we evaluate and track the value of the cost function (\ref{4A4}) with respect to the validation set, and select parameters when the cost function is minimized.
Finally, the performance of the proposed RNN model is verified with the test set, which is not used in the training stage.

Once every parameter of the proposed RNN model is adjusted appropriately, it is now ready to identify the channel condition based on the following simple hypothesis test:
{\setlength\arraycolsep{1pt}
\begin{eqnarray} \label{4A5}
&&H_0: h_\Theta(\mathbf{X})\geq \alpha,\ \mbox{LOS condition}\nonumber\\
&&H_1: h_\Theta(\mathbf{X})< \alpha,\ \mbox{NLOS condition},
\end{eqnarray}}%
where $H_0$ and $H_1$ are called null and alternative hypotheses, respectively, and $\alpha$ represents the decision threshold.
In this hypothesis test, the LOS detection rate or true positive rate (TPR) corresponds to the portion of correct decisions among all measurements under the LOS condition. Similarly, NLOS detection rate or true negative rate (TNR) corresponds to the portion of correct decisions among all measurements under NLOS condition. In addition, we define accuracy as the portion of correct decisions over all measurements.
These statistical values depend on the decision threshold $\alpha$, which is generally chosen to maximize a certain objective, for instance, the accuracy or the average of TPR and TNR.

\section{Performance Evaluation}

\subsection{Conventional Handcrafted Feature-Based Identifications}

For performance comparison, we refer to the two latest approaches introduced in~\cite{7130677, 7218588}, which exploit the CSI obtained by commodity WLAN devices to identify channel conditions.
The authors in \cite{7130677} first recover the CIR from the CSI, monitor the dominant path power over $P$ packet transmissions, and take skewness to the measured data to investigate the different variation patterns of the dominant path across the LOS and NLOS conditions.
Likewise, they track the standard deviation of the normalized CSI and take the kurtosis to the monitored data.
We denote these two methods as \emph{skewness-based} and \emph{kurtosis-based} identifications, respectively.

In a similar manner, authors in~\cite{7218588} extract phase information from CSI, apply the phase calibration method introduced in~\cite{monalisa}, and evaluate the phase variation for every sub-carrier.
They finally define a handcrafted feature, called $\rho$-factor, which is the average of phase variation weighted by the amplitude of the corresponding CSI. We denote this method as \emph{phase-based} identification.
As explained in Section III, our measurement system simultaneously acquires six spatial CSI sets.
Therefore, we can extract that number of handcrafted feature values for every $P$ packet transmission.
We take the median of six different values to generate a single representative value for each conventional feature.

Because on the asymmetric number of measured data under the LOS and NLOS conditions, the accuracy, which is simply defined by the portion of correct decisions, is not a desirable metric to evaluate performance in this paper.
Instead, we use the receiver operating characteristic (ROC) curve, which illustrates the relationship between TPR and false positive rate (FPR), which is defined by the portion of incorrect decisions among all measurements under the NLOS condition (i.e., FPR=1-TNR).
The ROC curve represents better performance as it approaches a point in the upper left edge, which implies a perfect decision.

Fig.~\ref{figRocConv} illustrates the ROC curve of conventional methods, where each handcrafted feature is extracted from $P=1000$ measurement data.
According to~\cite{6778037}, an electromagnetic wave can propagate around a human body and the body shadowing loss varies widely.
Therefore, in some cases, the receiver can still experience the effect of a dominant path, even though the direct path is shadowed by a human body.
For this reason, identifying the NLOS condition due to body shadowing is more challenging.
To address this problem, we represent the ROC curves of the conventional methods for two problems.
Fig.~\ref{figRocConv}(a) represents the identification between the LOS and NLOS conditions due to the building structure only.
In Fig.~\ref{figRocConv}(b), we consider every NLOS condition including human body shadowing.
The channel condition identification becomes more complicated if we consider the NLOS condition due to human body.
Each handcrafted feature is extracted from CSI data obtained by $P=1000$ packet transmissions.
In both cases, the skewness-based identification yields the best result among the existing schemes.

Unlike as claimed in the original work, the phase feature does not produce accurate results in this study.
Because this feature is valid for a stationary scenario, where the number of multi-path components remains unchanged during a measurement so that the phase feature captures the subtle variation in phase across different channel conditions.
However, our receiver continuously moves around to collect data and the multi-path components are frequently changed.
As a result, the phase of each OFDM sub-carrier fluctuates widely even though the phase calibration method proposed in~\cite{monalisa} is applied, and it is difficult to evaluate the slight variation in phase.

\begin{table}
\caption{Maximum average LOS/NLOS detection rate of conventional methods}
\label{tableAvg}
\centering
\begin{tabular}{|c|c|c|c|c|}
\hline
\# of packets, $P$         & Skewness~\cite{7130677}  &    Kurtosis~\cite{7130677} & Phase~\cite{7218588} \\\hline
10          & 60.5 \%   & 51.7 \% & 64.0 \%   \\\hline
50          & 70.0 \%   & 56.5 \% & 72.0 \%   \\\hline
100         & 75.2 \%   & 61.9 \% & 74.0 \%   \\\hline
500         & 82.9 \%   & 73.2 \% & 77.3 \%   \\\hline
1000        & 85.9 \%   & 73.2 \% & 78.3 \%   \\\hline
3000        & 91.5 \%   & 81.2 \% & 83.7 \%   \\\hline
\end{tabular}
\end{table}
%
%
%

Table I summarizes the maximum achievable average detection rate of existing schemes, i.e., the maximum (TPR+TNR)/2 value in the ROC curve.
The average detection rate using conventional handcrafted features tends to increase with measurement time.
The skewness-based method can achieve 91.5~\% of the average detection rate if the feature is extracted from $P$=3000 packet transmissions, and the accuracy using the same threshold level is measured to be 90.1~\%.
However, it takes quite a long time to achieve such level of accuracy, i.e., 30 s = 10~ms $\times$~3000.

\subsection{Performance of the Proposed RNN Model}

We train the proposed RNN model using the Tensorflow application running on a Windows environment.
We set the number of hidden layers as $D_h=10$ and use the Adam optimizer with the learning rate of $\eta=0.0005$~\cite{Adam}.
As mentioned in Section IV, we split the total measurement data into training, validation, and test  sets and prepare the input and output pairs from each set separately.
Also, the optimal parameter in the RNN model may vary depending on the environment, and the proposed model needs to be retrained if the propagation environment is significantly changed.
Every training process in this study takes much less than one hour using a workstation equipped with dual Intel(R) Xeon CPU E5-2680 v2 processors.

\begin{figure}[!t]
\centering
\includegraphics[width=0.48\textwidth]{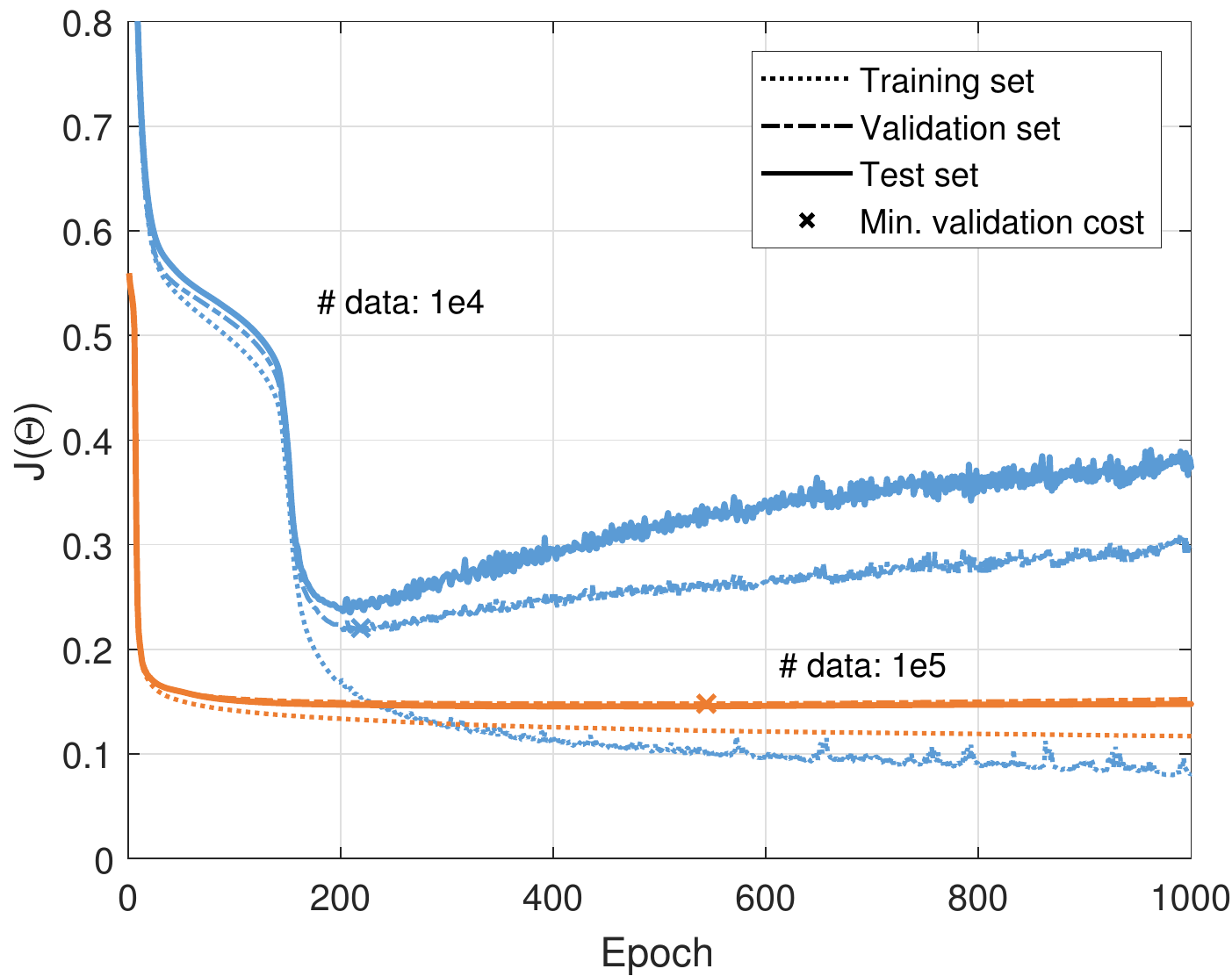}
\caption{Cost function with respect to training, validation, and test set. The minimum cost for validation set is marked as x.}
\label{figCost}
\end{figure}

\begin{table}
\caption{Performance depending on the number of data}
\label{tableAvg}
\centering
\begin{tabular}{|c|c|c|c|c|}
\hline
\# of data              & 5000  & 1e4 & 5e4 & 1e5  \\\hline
Minimum validation cost         & 0.2214  & 0.2133 & 0.170 & 0.165  \\\hline
Epochs to stop                  & 117     & 212    & 103    & 552  \\\hline
Decision threshold ($\alpha$)     & 0.103  & 0.097 & 0.176 & 0.125   \\\hline
Avg. detection rate     & 89.4 \%  & 92.2 \% & 92.9 \% & 93.7 \%  \\\hline
Accuracy                & 88.4 \% & 92.0 \% & 91.6 \% & 91.4 \%   \\\hline
\end{tabular}
\end{table}

At first, we verify the performance of the proposed RNN model depending on the number of data.
This number is expressed in an exponential expression (e.g., 1e4=10000) and includes training, validation, and test data.
We select certain amounts of data from these three data sets so that the portion of the selected data becomes 70~\%, 15~\%, and 15~\%, respectively.
Furthermore, we create each input data from $P=10$ packet transmissions and run the optimization algorithm up to 1000 iterations.
One iteration is also called epoch, which implies that the parameter of the proposed RNN model is updated using the entire training data.

Fig.~\ref{figCost} illustrates the variation in cost function with respect to training, validation, and test sets.
As the figure shows, the cost function for training data tends to decrease with the iteration, whereas those for validation and test data decrease to a certain iteration and then increase again.
In addition, the more data are used for training, the less is achieved cost can be achieved.
The minimum cost achieved for validation data is presented in the figure as x-mark.
The performance for different numbers of data is summarized in Table~II.
The fields denoted by \emph{Minimum validation cost} and \emph{Epochs to stop} imply the minimum cost function with respect to the validation set and the number of training epochs to achieve this cost, respectively.
The decision threshold is selected to maximize the average LOS/NLOS detection rate.

\begin{figure}[!t]
\centering
\includegraphics[width=0.47\textwidth]{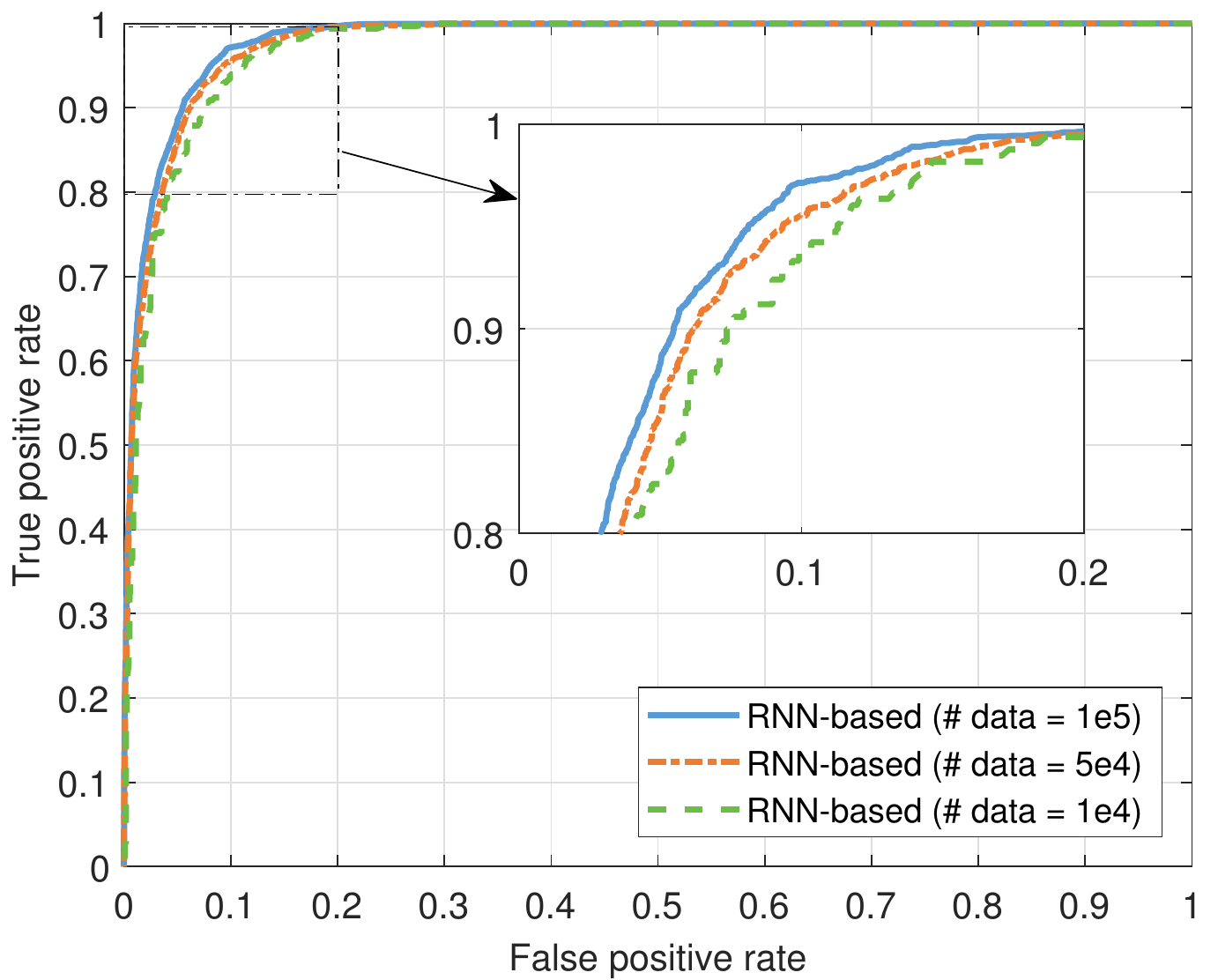}
\caption{ROC curves of LOS identification using the proposed RNN model depending on the number of data used; 70~\% of data is used for training purpose.}
\label{figRoc3}
\end{figure}

\begin{figure}[!t]
\centering
\includegraphics[width=0.47\textwidth]{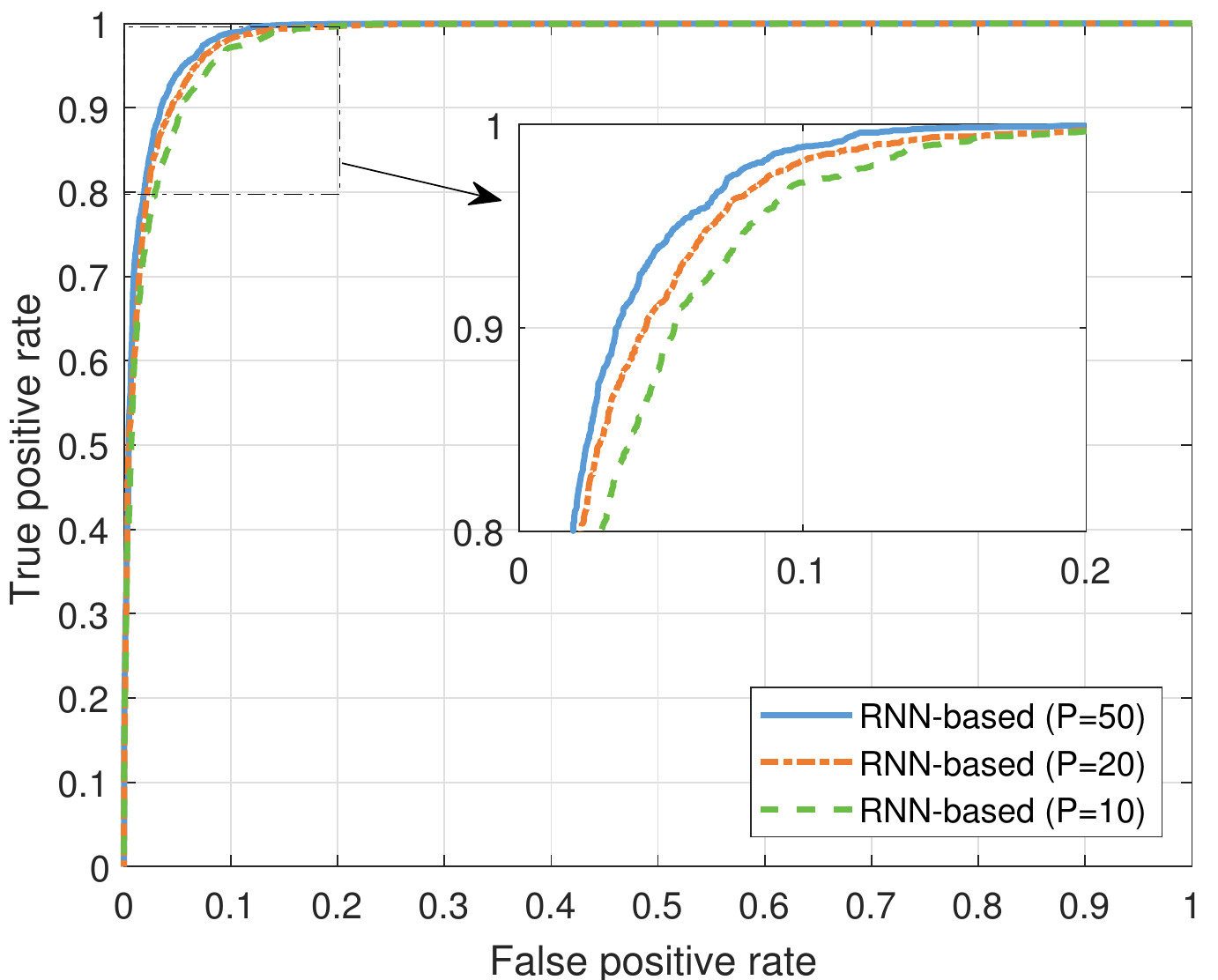}
\caption{ROC curves of LOS identification using the proposed RNN model depending on the number of measured packet transmissions.}
\label{figRocDh}
\end{figure}

The ROC curves of the proposed RNN model are depicted in Fig.~\ref{figRoc3}.
This figure indicates that the more number of data is used for training, the more accurate the results.
However, the performance gap between the cases using 5e4 data and 1e5 data is not large.
Regardless of the number of data used, the performance of the proposed LOS identification method exceeds the best conventional scheme depicted in Fig.~\ref{figRocConv}(b), even though the proposed method considers only $P=10$ packet transmissions.
Fig.~\ref{figRocDh} depicts the ROC curves of the proposed model depending on the number of packet transmissions.
The ROC curve approaches the perfect decision point as the number of monitored packet transmissions increases.

\subsection{Performance with CSI Only}

The reason why the proposed RNN model yields high accuracy even for the data observed in a very short time might be that the input data contain RSSI information.
As seen in Fig.~2, the range of RSSI under the LOS and NLOS channel conditions is separable, and thus, plays a major role in identifying the channel conditions.
Nevertheless, we also verify the performance of the RNN model considering only CSI data.
By doing so, we can compare the performance of the proposed and conventional schemes under the same condition.

\begin{figure}[!t]
\centering
\includegraphics[width=0.48\textwidth]{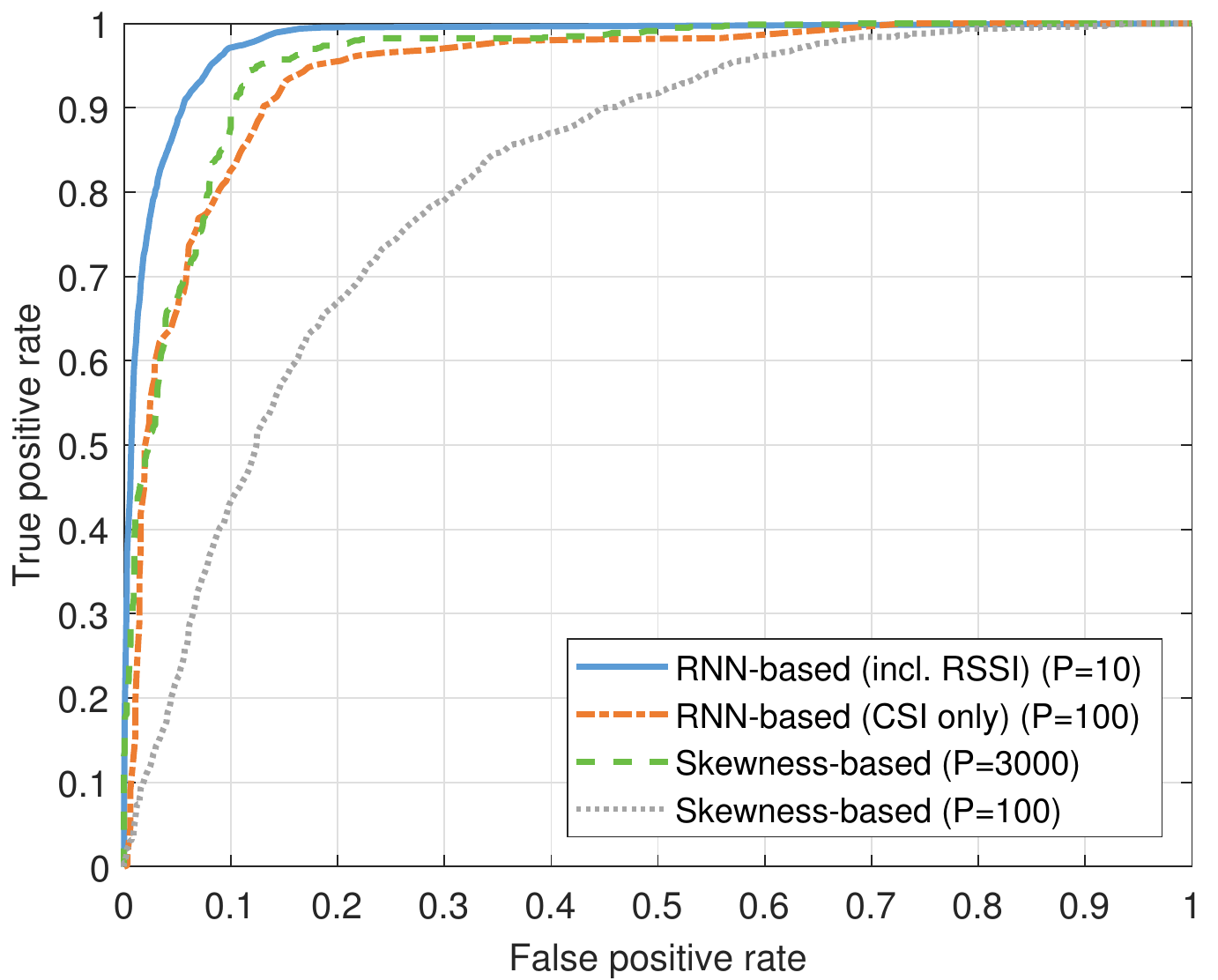}
\caption{ROC curves of LOS identification using the proposed RNN model and the conventional phase-based method.}
\label{figRocCompare}
\end{figure}

Fig.~\ref{figRocCompare} illustrates the ROC curves for the proposed RNN model and the best conventional method.
For the proposed model, we use two versions, where the first one considers RSSI with $P=10$ packet transmissions and the second one considers only CSI data obtained from more packet transmissions (i.e., $P=100$).
For the second case, we also take the median of six outputs of the spatial channel to increase accuracy, as we did for conventional schemes.
For the conventional scheme, we choose $P=100$ packet transmissions to consider the same number of packet transmissions with the proposed scheme and $P=3000$ to evaluate the performance when sufficient amount of data is observed.
Even though the proposee RNN model takes CSI data only, it can produce accurate results with $P=100$ packet transmissions, where the ROC curve is similar to that of the best conventional method with much more packet transmissions.

\begin{figure}[!t]
\centering
\includegraphics[width=0.48\textwidth]{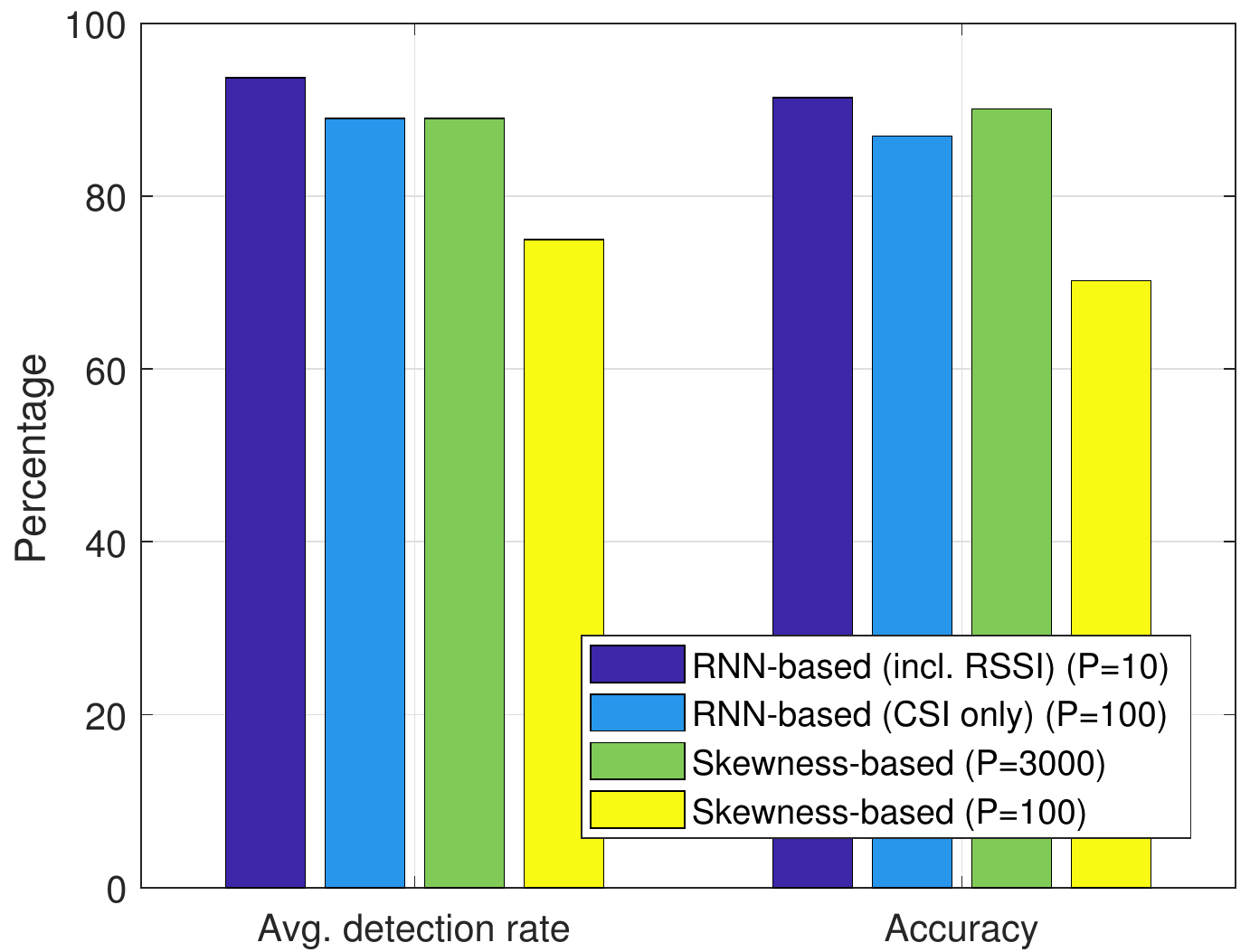}
\caption{Maximum achievable average detection rate and accuracy for the proposed RNN model and the best conventional model.}
\label{figBar}
\end{figure}

Fig.~\ref{figBar} depicts the average detection rate and accuracy of channel condition identification.
We choose an optimal decision threshold to maximize the average detection rate.
As we can see, the proposed RNN model with RSSI information instantly yields the best result, where the maximum achievable average detection rate and accuracy are 93.7~\% and 91.4~\%, respectively.
On the other hand, the best conventional method monitoring the variation in the dominant power produces a meaningful figure for data obtained for quite a long time.

\section{Conclusion}

In this paper, we proposed a channel condition identification method using an RNN structure with an LSTM block, which takes a series of CSI.
Instead of utilizing handcrafted features to identify the channel conditions, the proposed model efficiently learns a good representation between raw input and output data during the training data.
Moreover, it combined cross-layer information (e.g., CSI and RSSI) appropriately to produce accurate results.
For training, we collected excessive measurement data under an indoor office environment using commodity WLAN device, and used early stopping method to avoid the risk of over-fitting.
The proposed method yielded high accuracy even for data acquired in a very short time.

\ifCLASSOPTIONcaptionsoff
  \newpage
\fi



%

\begin{IEEEbiography}[{\includegraphics[width=1in,height=1.25in,clip,keepaspectratio]{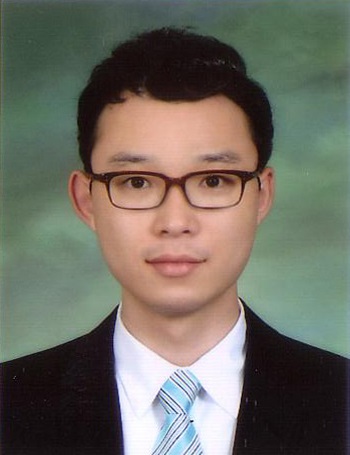}}]{Jeong-Sik Choi}
received the B.S. degree in electrical engineering  in 2010 from Pohang University of Science and Technology (POSTECH), Pohang, Korea, and the M.S. and Ph. D. degrees in electrical engineering from Seoul National University, Seoul, Korea, in 2012 and 2016, respectively.
From 2016 to 2017, he was a Senior Researcher with the Institute of New Media and Communications (INMC), Seoul National University, Seoul, Korea.
He is now a Postdoctoral Researcher at Intel Labs, Intel Corporation, Santa Clara, CA, USA.
His research interests include wireless propagation channel measurement \& modeling, wireless resource management, and application of machine learning algorithms for indoor and outdoor positioning techniques.
\end{IEEEbiography}

\begin{IEEEbiography}[{\includegraphics[width=1in,height=1.25in,clip,keepaspectratio]{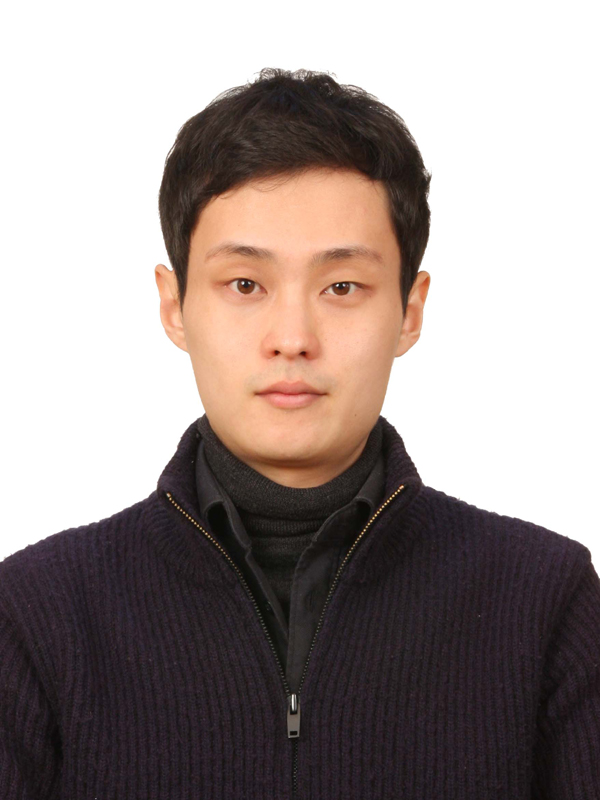}}]{Woong-Hee Lee}
received the B.S. degree in electrical engineering in 2009 from Korea Advanced Institute of Science and Technology (KAIST), Daejeon, Korea, and the Ph. D. degree in electrical engineering in 2017 from Seoul National University, Seoul, Korea. From 2017, he is with the Advanced Standard Research and Development Laboratory, LG Electronics, Seoul, Korea. His research interests include resource management and game theoretical analysis of wireless networks.
\end{IEEEbiography}

\begin{IEEEbiography}[{\includegraphics[width=1in,height=1.25in,clip,keepaspectratio]{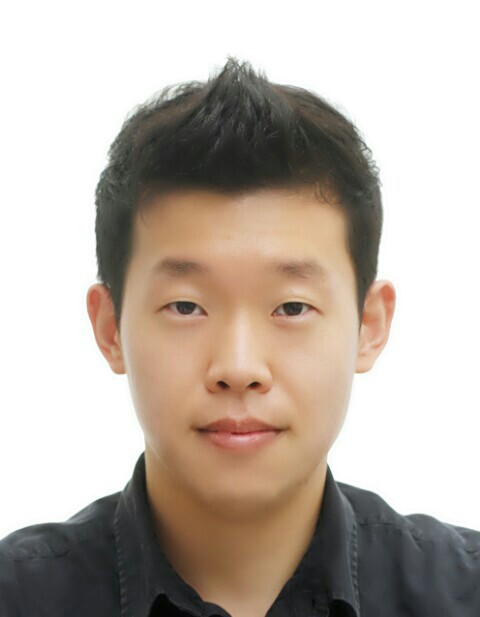}}]{Jae-Hyun Lee}
received the B.S. degree in electrical engineering in 2012 from Pohang University of Science and Technology (POSTECH), Pohang, Korea. He entered the combined master’s and doctoral program in 2012 and is currently pursuing his Ph.D. in electrical engineering from Seoul National University, Seoul, Korea. His research interests include millimeter wave communication, wireless channel and ray-tracing simulation of wireless networks.
\end{IEEEbiography}

\begin{IEEEbiography}[{\includegraphics[width=1in,height=1.25in,clip,keepaspectratio]{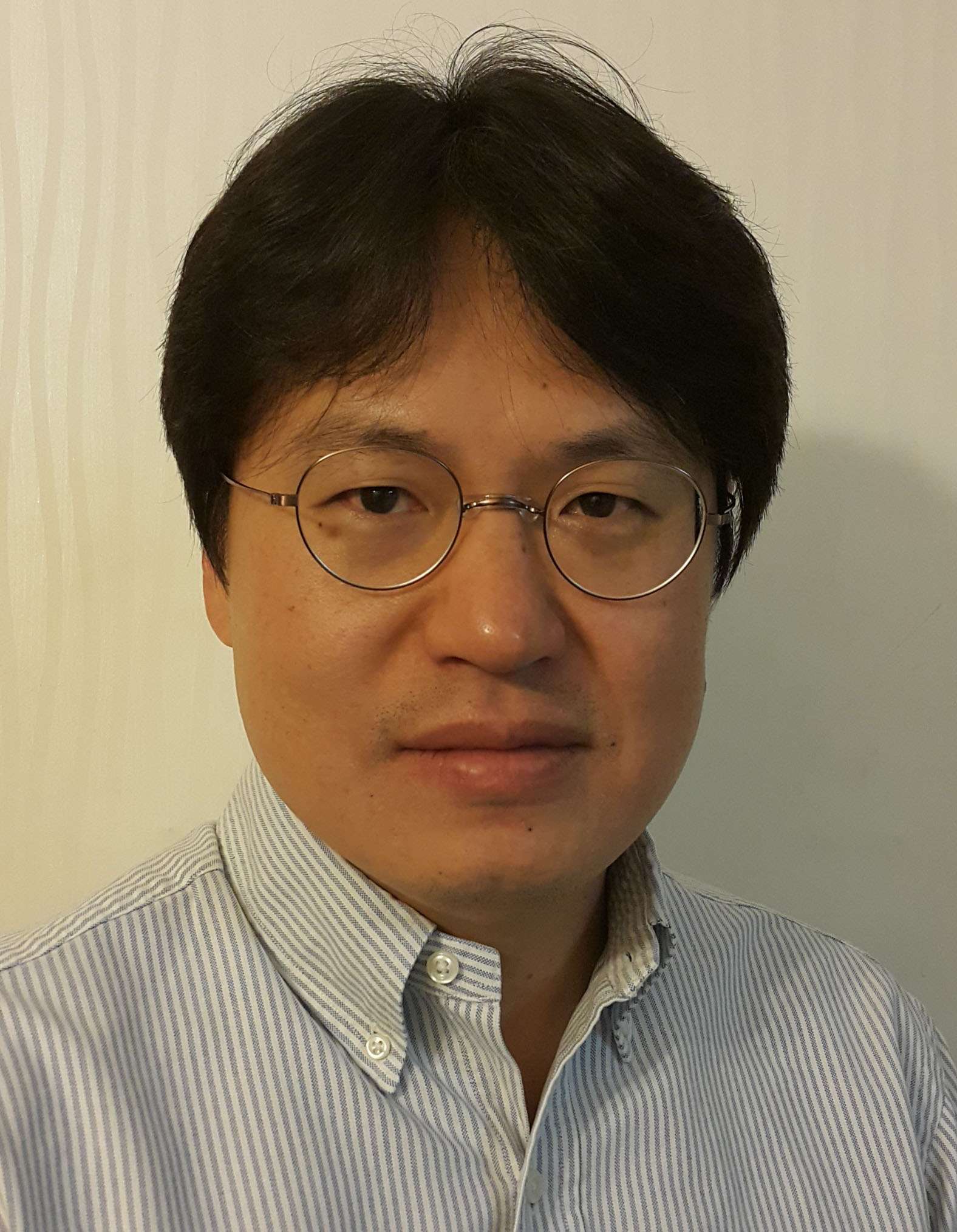}}]{Jong-Ho Lee}
(M'13) received the B.S. degree in electrical engineering and the M.S. and Ph.D. degrees in electrical engineering and computer science from Seoul National University, Seoul, Korea, in 1999, 2001, and 2006, respectively. From 2006 to 2008, he was a Senior Engineer with Samsung Electronics, Suwon, Korea. From 2008 to 2009, he was a Postdoctoral Researcher with the Georgia Institute of Technology, Atlanta, GA, USA. From 2009 to 2012, he was an Assistant Professor with the Division of Electrical Electronic and Control Engineering, Kongju National University, Cheonan, Korea. Since 2012, he has been with the faculty of the Department of Electronic Engineering, Gachon University, Seongnam, Korea. His research interests are in the areas of wireless communication systems and signal processing for communication with current emphasis on multiple-antenna techniques, multi-hop relay networks, physical layer security, and full-duplex wireless communications.
\end{IEEEbiography}

\begin{IEEEbiography}[{\includegraphics[width=1in,height=1.25in,clip,keepaspectratio]{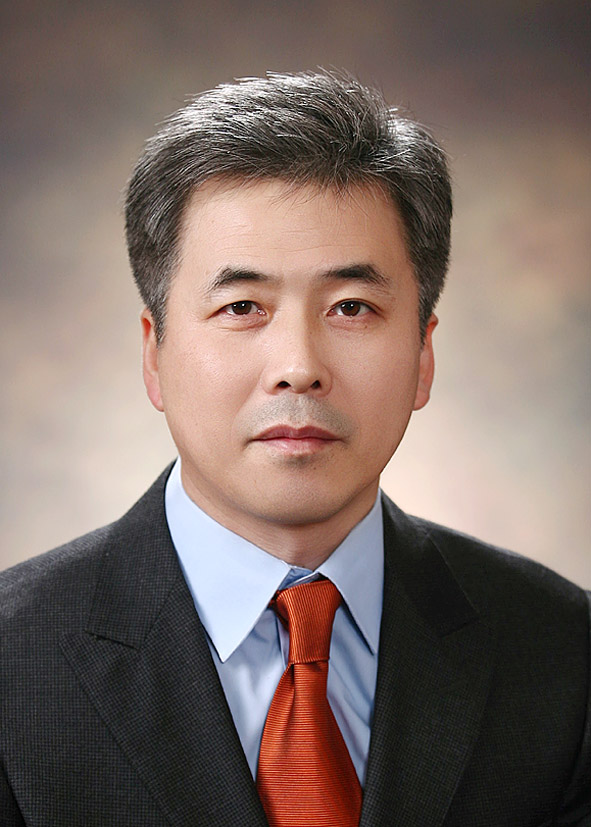}}]{Seong-Cheol Kim}
(S'91-M'96-SM'12) received the B.S. and M.S. degrees in electrical engineering from Seoul National University, Seoul, Korea, in 1984 and 1987, respectively, and the Ph.D. degree in electrical engineering from Polytechnic Institute of NYU, Brooklyn, NY, in 1995. From 1995 to 1999, he was with the Wireless Communications Systems Engineering Department, AT\&T Bell Laboratories, Holmdel, NJ. Since 1999, he has been a Professor with the Department of Electrical Engineering and Computer Science, Seoul National University, Seoul, Korea. His research area covers systems engineering of wireless communication including propagation channel modeling, localization, and resource management. Currently, he is also interested in power-line communication and automotive radar system.
\end{IEEEbiography}

%
%

%

%






\end{document}